\documentstyle[11pt,aaspp4,flushrt,tighten]{article}
\singlespace
\def\kms{\,{\rm km\,s^{-1}}}
\def\kmsmpc{\,{\rm km\,s^{-1}\,Mpc^{-1}}}

\def\lunits{\,{\rm ergs\,s^{-1}}}

\def\sfr{\,{\rm M_\odot\,yr^{-1}}}
\def\page{\vfill\eject}

\def\etal{{et al.\ }}

\def\spose#1{\hbox to 0pt{#1\hss}}
\def\lta{\mathrel{\spose{\lower 3pt\hbox{$\mathchar"218$}}
     \raise 2.0pt\hbox{$\mathchar"13C$}}}
\def\gta{\mathrel{\spose{\lower 3pt\hbox{$\mathchar"218$}}
     \raise 2.0pt\hbox{$\mathchar"13E$}}}

\begin{document}
\title{Compton Echoes from Gamma-ray Bursts}

\author{Piero Madau\altaffilmark{1,2}, Roger D. Blandford\altaffilmark{1,2,3}, 
and Martin J. Rees\altaffilmark{1}}
\altaffiltext{1}{Institute of Astronomy, Madingley Road, 
Cambridge CB3 0HA, UK.}
\altaffiltext{2}{Institute for Theoretical Physics, University of California,
Santa Barbara, CA 93106--4030, USA.}
\altaffiltext{3}{Theoretical Astrophysics, 130-33 Caltech, Pasadena, CA 91125, 
USA.}

\begin{abstract}
\noindent 
Recent observations of gamma-ray bursts (GRBs) have provided growing evidence 
for collimated outflows and emission, and strengthened the connection between 
GRBs and supernovae. If massive stars are the progenitors of GRBs, the 
hard photon pulse will propagate in the pre-burst, dense environment. 
Circumstellar material will Compton scatter the prompt GRB radiation and 
give rise to a reflection echo. We calculate luminosities, spectra, and light
curves of such Compton echoes in a variety of emission geometries and ambient
gas distributions, and show that the delayed hard X-ray flash from a pulse 
propagating into a red supergiant wind could be detectable by {\it Swift} out 
to $z\sim 0.2$.  Independently of the $\gamma$-ray spectrum of the prompt 
burst, reflection echoes will typically show a high-energy cutoff between 
$m_ec^2/2$ and $m_ec^2$ because of Compton downscattering. At fixed burst
energy per steradian, the luminosity of the reflected echo is proportional to 
the beaming solid angle, $\Omega_b$, of the prompt pulse, while the number of 
bright echoes detectable in the sky above a fixed limiting flux increases as 
$\Omega_b^{1/2}$, i.e. it is smaller in the case of more collimated jets.
The lack of an X-ray echo at $\sim $ one month delay from 
the explosion poses severe constraints on the possible existence of a lateral 
GRB jet in SN 1987A. The late $r$-band afterglow observed in GRB990123 is
fainter than the optical echo expected in a dense red supergiant environment 
from a isotropic prompt optical flash. Significant MeV delayed emission may be 
produced through the bulk Compton (or Compton drag) effect resulting from the 
interaction of the decelerating fireball with the scattered X-ray radiation.

\end{abstract} 

\section{Introduction}
The nature of the progenitors of gamma-ray bursts (GRBs) remains an unsettled
issue after more than three decades of research (e.g. M\'{e}sz\'{a}ros, Rees, \&
Wijers 1999; Paczy\'{n}ski 1999; Piran 1999). The discovery of X-ray (Costa
\etal 1997) and optical (van Paradijs \etal 1997) afterglows has provided 
the breakthrough needed to establish the cosmological nature of these events.
While the implied huge energy release per steradian must produce a 
relativistically 
expanding fireball (e.g. M\'{e}sz\'{a}ros \& Rees 1992), it is not yet clear 
if this expansion is quasi-spherical or highly collimated, or if the degree of 
beaming differs between the prompt GRB and the delayed emission. The observed 
distribution of optical afterglows with respect to their host galaxies may 
suggest that some GRBs are associated with star forming regions, 
and therefore with the explosions of massive stars rather than with merging 
neutron stars (e.g. Paczy\'{n}ski 1998; Fruchter \etal 1999). Recent 
observations of a spatial and temporal coincidence between the supernova (SN) 
1998bw and GRB980425 (Galama \etal 1998) have added support to the idea that 
at least some GRBs may be related to some type of supernova explosion. 
The observation of a red component,
a factor of 60 higher in flux than the extrapolated afterglow in GRB980326, 
has also been explained with a supernova, and, if true, it would strengthen
the connection of GRBs with massive stars (Bloom \etal 1999; see also Reichart
1999 for similar evidence in GRB970228; but see Esin \& Blandford 2000 for an 
alternative explanation involving dust scattering). 
If GRBs are associated with SN-like events, energetic considerations suggest 
that some fireballs be collimated into a 
solid angle $\Omega_b\ll 4\pi$, i.e. GRBs involve strongly 
asymmetric outflows. Without beaming the inferred (`isotropic equivalent') 
energy of GRB990123, $\sim 4\times 10^{54}\,$ ergs (Kulkarni \etal 1999), 
rules out  stellar models. Additional circumstantial evidence for 
jet-like bursts comes from the decline (`beaming break') observed in some 
afterglow light curves (Kulkarni \etal 1999; Harrison \etal 1999) and 
attributed to the sideway expansion of non-spherical ejecta, although some
difficulties remain (Moderski, Sikora, \& Bulik 2000).

It is an unavoidable consequence of a massive star progenitor model for GRBs 
that the hard emission will propagate in a dense circumstellar environment, 
such as a pre-burst stellar wind (Chevalier \& Li 1999). 
Shells of significantly enhanced gas density may also exist in the immediate 
neighborhood of a GRB. This would be the case, for example, in scenarios where 
a supernova occurs shortly before the burst (like in the delayed collapse of a 
rotationally-stabilized neutron star, Vietri \& Stella 1998) so that the 
metal-enriched supernova remnant shell does not have time to disperse. The
deceleration of a pre-supernova wind by the pressure of the surrounding 
medium could also create circumstellar shells, as would the interaction of fast
and slow winds from massive stars (as observed in the case of SN 1987A). The 
ambient material will then efficiently scatter the prompt GRB radiation and, 
because of light travel time effects, produce a luminous (albeit unresolvable) 
reflection echo. The detection of scattered light may provide unique 
information on the environment of GRBs and on their emission properties. 
Conversely, one may use the lack of evidence for a Compton echo to set 
constraints on the density of circumstellar material and the burst energetics. 
In this Paper we compute the expected properties of such Compton flashes in a 
variety of scenarios, and show that the delayed hard X-ray emission 
associated with the echo reflected by a red supergiant wind is significant and 
could be detectable by {\it Swift} out to a redshift of $z\sim 0.2$.

\section{Compton echoes from circumstellar gas}

For simplicity,  we will approximate the prompt GRB as a collimated photon 
pulse that maintains a constant luminosity for a time $\Delta$. It is commonly 
proposed that the burst is produced when the kinetic 
energy of a relativistically 
expanding fireball is dissipated at a radius $r_{\rm sh}\sim 10^{13.5}\,$cm 
due to, e.g., internal shocks, and radiated away as
$\gamma$-rays through synchrotron emission. The scattered light is observed at 
a time $t\gg \Delta$ ($t$ is measured since the burst is first detected), 
when the radiation beam can be visualized as a shell of radius $ct\gg r_{\rm 
sh}$ and thickness $c\Delta $.
The (equal-arrival time) scattering material lies on the paraboloid having the 
burst at its focus and its axis along the line of sight,
\begin{equation}
r={ct\over 1-\cos\theta},
\end{equation} 
where $\theta$ is the angle between the line of sight and the direction of the 
reflecting gas as seen by the burst (e.g. Blandford \& Rees 1972). 

The sudden brilliance of a GRB will be reflected by the circumstellar gas 
to create a `Compton echo'. If $E$ is the total energy, 
$E\equiv\int E_{\epsilon\Omega}d\epsilon d\Omega$, emitted by the 
burst ($E_{\epsilon\Omega}$ is the energy emitted per unit energy 
$\epsilon$ and unit solid angle $\Omega$ along the $\theta$ direction, 
$E_{\epsilon\Omega}=E_{\epsilon} /4\pi$ in the case of 
isotropic emission), 
then the equivalent isotropic luminosity (as inferred by a distant observer) 
of the echo reflected at a distance $r$ from the site of the burst is
\begin{equation}
L_{\epsilon'}=4\pi\int n_e(r,\theta) E_{\epsilon\Omega} {d\sigma\over 
d\Omega} {dr\over dt}d\Omega,  
\label{eq:lum}
\end{equation}
where the integral is over the beaming solid angle $\Omega_b$ of the prompt
pulse, $r$ is given by equation (1), 
\begin{equation}
{d\sigma\over d\Omega}={3\sigma_T\over 16\pi} 
\left({\epsilon'\over \epsilon}\right)^2
\left({\epsilon\over \epsilon'}+{\epsilon'\over \epsilon}-\sin^2\theta\right)
\label{eq:kn}
\end{equation} 
is the differential Klein-Nishina cross section for unpolarized incident 
radiation (e.g. Rybicki \& Lightman 1979), $\sigma_T$ is the Thomson cross 
section, 
\begin{equation}
\epsilon'=\epsilon \left[1+{\epsilon\over m_ec^2} (1-\cos\theta)\right]^{-1}
\end{equation}
is the energy of the scattered photon, and $n_e$ the local electron density.
Equation (\ref{eq:lum}) assumes that photons scatter only once and that 
absorption can be neglected (which is a good approximation at observed 
energies $\gta 10\,$keV). 

In the relativistic regime, photons experience a reduced cross section and 
scatter preferentially in the forward direction. Electron recoil  further 
suppresses the reflected echo at $\gamma$-ray energies. In the non-relativistic
regime, $\epsilon\approx \epsilon'$, equation (\ref{eq:kn}) reduces to the 
classical Thomson limit, and the observed echo luminosity becomes  
\begin{equation}
L_\epsilon={3\over 4} \int n_e E_{\epsilon\Omega} \sigma_T c\, {1+\cos^2\theta 
\over 1-\cos\theta}d\Omega \label{eq:Thom}.
\end{equation}
Below we discuss a few scenarios in the Thomson regime which well illustrate 
the range of possible emission geometries and ambient gas distributions.

\subsection{Uniform medium extending to ${\bf r=R}$}
\noindent {\it a) Isotropic burst.}\quad In this case the echo is 
dominated by gas along the line of sight, and 
\begin{equation}
L_\epsilon={3\over 4} n_e E_\epsilon\sigma_T c \left[\ln({2R\over ct})
-\left(1-{ct\over 2R}\right)^2\right]
\end{equation}
for $t<2R/c$. The emission diverges logarithmically at zero lag and then 
decreases monotonically to zero at $t=2R/c$.

\noindent {\it b) Collimated burst.}\quad Here the zone of emission 
propagates out along the approaching and receding jets (assumed to have
equal energy $E_\epsilon/2$) until $r=R$, and 
\begin{equation}
L_\epsilon={3\over 4} n_e E_\epsilon\sigma_T c\,\left\{\begin{array}{ll} 
{1+\cos^2\theta \over 1-\cos^2\theta}; &
\mbox{~~~~~~~~$0<t<R(1-\cos\theta)/c$,} \\ {1+\cos^2\theta \over 
2(1+\cos\theta)}; &
\mbox{~~~~~~~~$R(1-\cos\theta)/c<t<R(1+\cos\theta)/c$,} 
\end{array}
\right. 
\end{equation}
where $\theta$ is the angle between the line of sight and the approaching beam.

\subsection{Constant velocity wind with ${\bf n_e=Ar^{-2}}$}
\noindent {\it a) Isotropic burst.}\quad The reflected flash comes from 
the surface of the paraboloid and is dominated by the apex behind the source,
\begin{equation}
L_\epsilon={A E_\epsilon \sigma_T \over ct^2}. \label{eq:Liso}
\end{equation}
The echo declines with time faster than most observed afterglows.

\noindent {\it b) Collimated burst.}\quad  In this case  
\begin{equation}
L_\epsilon={3 A E_\epsilon \sigma_T\over 4ct^2}\, (1+\cos^2\theta), 
\end{equation}
and the echo is dominated by the receding jet as, at a given observer time, 
it originates closer to the GRB where the density is greater.

\subsection{Thin spherical shell of radius ${\bf R}$ and column ${\bf N_e}$}
This scenario may arise if the progenitor star loses most of its mass quite 
rapidly a short but finite time prior to the explosion. 

\noindent {\it a) Isotropic burst.}\quad One derives 
\begin{equation}
L_\epsilon={3 N_e E_\epsilon \sigma_T c\over 4R}\left(1-{ct\over R}+
{c^2t^2\over 2R^2}\right) \label{eq:shell}
\end{equation}
for $ct<2R$. As the paraboloid sweeps up the shell, the echo luminosity 
decreases, reaches a minimum when $t=R/c$, and increases again till the back 
of the shell is passed by the paraboloid. 

\noindent {\it b) Collimated burst.}\quad In this case the echo light curve
is the sum of two delta functions,
\begin{equation}
L_\epsilon={3\over 8} N_e E_\epsilon \sigma_T (1+\cos^2\theta)\{\delta[t-
R(1-\cos\theta)/c]+\delta[t-R(1+\cos\theta)/c]\}.
\end{equation}
The two spikes of emission are seen separated by an interval $2R\cos\theta/c$.

\subsection{Slender annular ring of radius ${\bf R}$ and angular width 
${\bf \Delta\theta}$}
This model is inspired by observations of SN 1987A and $\eta$Carinae which both
exhibit dense equatorial rings. Let the ring have an inclination $i$.
The echo associated with an isotropic burst has then luminosity
\begin{equation}
L_\epsilon={3 N_e E_\epsilon \Delta\theta \sigma_T c\over 4\pi R}
{\left(1-{ct\over R}+{c^2t^2\over 2R^2}\right)\over
\left({2ct\over R}-{c^2t^2\over R^2}-\cos^2i\right)^{1/2}};~~~~~~~~~
R(1-\sin i)/c<t<R(1+\sin i)/c.
\end{equation}
The emission is maximized at the beginning and at the end of the response.
The echo temporal behaviour in some of the scenarios discussed
above is shown in Figure 1. 

\section{Wind models}
 
In the following we will assume a broken power-law for the `typical' GRB 
spectrum,
\begin{equation}
\epsilon E_\epsilon\propto \left\{ \begin{array}{ll} \epsilon & \mbox{if 
$\epsilon\le 250\,$ keV,} \\ \epsilon^{-0.25} & \mbox{if $\epsilon>250\,$ keV.} \end{array}
\right.\label{eq:spe}
\end{equation}
This is consistent with a recent analysis of $\sim 150$ spectra obtained 
by the Burst And Transient Source Experiment (BATSE) on the {\it Compton 
Gamma Ray Observatory} (Preece \etal 2000). No attempt has been made to correct
the observed break energy for the mean redshift of the GRB population.

\subsection{Radiative acceleration of ambient material}

Equation 
(\ref{eq:lum}) only applies to a scattering medium which is either at rest
or moving with sub-relativistic speed. As shown by Madau \& Thompson (2000) 
and Thompson \& Madau (2000), a strong burst of radiation will have important 
dynamical effects on the surrounding ISM. Optically thin material overtaken by 
an expanding photon shell at radius $r$ will develop a large bulk Lorentz 
factor $\Gamma$ when the energy deposited by Compton 
scattering exceeds the rest-mass energy of the scatterers, i.e. when 
\begin{equation}
{\cal S}\equiv {E\bar \sigma\over \Omega_b r^2 \mu_e m_p c^2}\gg 1, \label{eq:R}
\end{equation} 
where $\Omega_b$ is the burst beaming angle, $\mu_e$ is the molecular weight 
per electron, $m_p$ the proton mass, and $\bar \sigma$ is the spectrum-weighted
total cross section,
\begin{equation}
\bar \sigma=E^{-1}\int E_\epsilon {d\sigma\over d\Omega} d\Omega d\epsilon
\approx 0.2\,\sigma_T
\end{equation} 
from equations (\ref{eq:kn}) and (\ref{eq:spe}). 
In the case of a pulse of `isotropic-equivalent' energy $E (4\pi/\Omega_b)=
10^{53}\,$ergs, propagating into a medium with $\mu_e=2$, one can then define 
a characteristic distance 
\begin{equation}
r_c({\cal S}\sim 1)\approx 6\times 10^{14}\,{\rm cm} 
\left({ 4\pi E\over 10^{53}\,\Omega_b\, {\rm ergs}}\right)^{1/2}, \label{eq:rc}
\end{equation}
such that for $r\ll r_c$ dynamical effects become important and may suppress  
the scattering rate (by a factor $\sim 1-\beta$).

More quantitatively, consider a plane-parallel photon pulse  of thickness 
$c\Delta $ propagating into the circumstellar gas. A parcel of matter moving 
radially with speed $c\beta=dr/dt$ will be accelerated 
at the rate  
\begin{equation}
{d\Gamma \over dr}=\Gamma^2(1-\beta)^2 {{\widetilde{\cal S}}\over c\Delta } 
\label{eq:dGam}
\end{equation}
where now ${\widetilde{\cal S}}$ includes a correction to $\bar \sigma$ as
particles moving at relativistic speed scatter an increasing fraction of
(redshifted) photons with the full Thomson cross section (i.e. $\bar \sigma
\rightarrow \sigma_T$). Matter initially at radius $r$ and accelerated from 
rest will surf the photon shell over a distance $\Delta r$ such that 
\begin{equation}
c\Delta=\int_r^{r+\Delta r} dr\, {1-\beta\over \beta}. \label{eq:Dr}
\end{equation}
The radiative force vanishes when the photon shell moves past the particle at 
$r+\Delta r$.
When $\Delta r\ll r$, the inverse square dilution of flux can be neglected,
and equation (\ref{eq:Dr}) can be rewritten using (\ref{eq:dGam}) as
\begin{equation}
\widetilde{\cal S}=\int_1^{\Gamma_{\rm max}} {d\Gamma\over 
\Gamma^2\beta(1-\beta)}. 
\end{equation}
This can be integrated exactly to yield a maximum Lorentz factor
\begin{equation}
\Gamma_{\rm max}=\cosh[\ln(1+\widetilde{\cal S})] 
\label{eq:Gam}
\end{equation}
which decreases with distance from the source. In the relativistic limit 
$\Gamma_{\rm max}\rightarrow 
{\widetilde{\cal S}}/2$ (Madau \& Thompson 2000). The acceleration distance is 
\begin{equation}
\Delta r ={c\Delta\over \widetilde{\cal S}}\int_1^{\Gamma_{\rm max}} 
{d\Gamma\over \Gamma^2(1-\beta)^2} 
=c\Delta {\widetilde{\cal S}\over 2}\left(1+{\widetilde{\cal S}\over 
3}\right).
\end{equation}
Figure 2 shows the bulk Lorentz factor derived from a numerical integration
-- including Klein-Nishina corrections --  
of equation (\ref{eq:dGam}), for a burst of isotropic-equivalent energy 
$10^{53}\,$ergs, duration $\Delta=10\,$s, and spectrum as in (\ref{eq:spe}). 
As expected, the acceleration at large radii takes place on a distance
$\Delta r\ll r/c$, and the outflow becomes sub-relativistic at about $r_c$. 
Two effects must be noted here: (1) The accelerated medium will be compressed
into a shell of thickness $\sim r/\Gamma_{\rm max}^2$. Shocks may form when 
inner shells (which move faster and are more compressed) run into outer 
shells, and material will accumulate at $r_c$. In a $r^{-2}$ density profile, 
the electron scattering optical depth in the wind from 
$r_c$ to infinity is $\tau_c=n_e(r_c)\bar\sigma r_c$. The mass accumulated
at $r_c$ by a burst of energy $E_\Omega$ per steradian is 
\begin{equation}
{E_\Omega\tau_c\over c^2}=0.006\,{\rm M_\odot~sr^{-1}}\,\left({E_\Omega\over
10^{52}\, {\rm ergs~sr^{-1}}}\right)\,\tau_c.
\end{equation}
Pre-acceleration of the ambient medium by the prompt radiation pulse will slow 
down the deceleration of the fireball ejecta (Thompson \& Madau 2000);  
and (2) We have solved the 
equations above assuming the ambient medium to be composed of a baryonic
plasma. It has been shown by Thompson \& Madau (2000) that $e^+e^-$ 
pair creation occurs in GRBs when side-scattered photons collide with the 
main $\gamma$-ray beam, an effect which amplifies the density of scattering 
charges in the ambient medium. The pair density will exponentiate when the 
photon shell is optically thick to photon collisions, i.e. when 
$\tau_{\gamma\gamma}\approx n_\gamma \sigma_T c\Delta/4\sim 1$. As
$\tau_{\gamma\gamma}\approx {\cal S} \mu_e m_p c^2/\langle \epsilon\rangle$, 
runaway pair production may occur well beyond  the radius $r_c$ defined in 
(\ref{eq:rc}). When pairs are produced in sufficient numbers, i.e. when 
$2m_en_{e^+}\gg m_pn_p$, the mean mass per scattering charge drops to 
$\sim m_e$. Because of the reduced inertia per particle, and also because
pair-producing collisions impart direct momentum to the gas, such a pair-loaded 
plasma may, under some circumstances,  be more efficiently accelerated to 
relativistic bulk velocities than a baryonic gas. This could 
increase the value of $r_c$ in equation (\ref{eq:rc}) by as much as a factor
$(\mu_e m_p/m_e)^{1/2}$. On the other hand, runaway pair creation will also 
boost the scattering optical depth of circumstellar material, thus producing 
brighter echoes at later times. For simplicity, in the rest of this paper 
we will limit our discussion to a {\it baryonic} scattering medium at 
$r\gta r_c$, where dynamical effects can be neglected. We defer a proper 
treatment of reflected echoes in a $e^+e^-$ pair-dominated wind to another work.

\subsection{Red supergiant winds}

In the case of a massive progenitor scenario, such as a `collapsar' (MacFadyen
\& Woosley 1999) or `hypernova' (Paczy\'{n}ski 1998), it is known that red 
supergiants and Wolf-Rayet (WR) stars have strong winds. If the progenitor 
is a Type 1b or Type 1c SN then it must have lost its hydrogen and
perhaps its helium envelope at some earlier time. The winds
from typical red supergiants are slow-moving and dense, with   
with velocities $v_w\approx 10-20\,\kms$ and mass loss rates between 
$10^{-6}$ and $10^{-4}\,\sfr$.   

As a `representative' red supergiant wind consider the case of SN 
1993J. While in a steady, spherically symmetric wind the electron density 
drops as 
\begin{equation}
n_e(r)={\dot M\over 4\pi v_w r^2 \mu_e m_p},
\end{equation}
deviations from a $r^{-2}$ density gradient towards a flatter slope, $n_e
\propto r^{-1.5}$, have been inferred in the circumstellar medium of this
supernova by Fransson, Lundquist, \& Chevalier 
(1996), and are possibly caused by a variation of the mass-loss rate from 
the progenitor or by a non-spherical geometry. Following Fransson \etal 
one can write  
\begin{equation}
n_e(r)\approx 10^8\,{\rm cm}^{-3} \left({r\over 10^{15}\,{\rm cm}}\right)
^{-1.5}\left({\dot M\over 
4\times 10^{-5}\,\sfr}\right) \left({v_w\over 10\,\kms}\right)^{-1} 
\label{eq:ne93J}
\end{equation} 
up to $r=2\times 10^{16}\,$cm, while at larger radii the observations appear
to be consistent with a $r^{-2}$ law. 

Figure 3 shows the reflected echo at $t=8, 24$, and 72 hours of a 
two-sided GRB jet inclined at an angle $\theta$ to the line of sight,
propagating through a SN 1993J-like dense environment.
The prompt pulse was assumed to radiate a total of $10^{52}\,$ ergs with the 
spectrum given in equation (\ref{eq:spe}), each jet having equal strength and
being invisible to the observer.
It is instructive to look at the relative contribution of the approaching and 
receding beams. On the 
equal-arrival time paraboloid, the receding beam is reflected by gas that is
closer to the source and denser: its contribution dominates the echo at all
energies where scattering occurs in the Thomson regime. Above 150 keV, 
however, recoil can no longer be neglected, and it is the approaching beam 
(whose photons are seen after small-angle scattering) which dominates the 
reflected flash at high energies.  
The total spectral energy distribution therefore mirrors the prompt burst at 
low energies, but is much steeper beyond a few hundred keV. In the limit
$\theta=90^\circ$ both beams are detected in reflected light after wide-angle
scattering, and the echo is suppressed above 511 keV by Compton downscattering
as $\epsilon'\rightarrow m_ec^2/(1-\cos\theta)$ for $\epsilon\gg m_ec^2$.
The scattered luminosity in Figure 3
drops initially as $t^{-1.5}$ for $ct<2\times 10^{16} (1-\cos\theta)\,$ cm, 
to steepen to $t^{-2}$ first at $\gamma$-ray energies (when the approaching beam
encounters the $r^{-2}$ density profile), and only later -- depending on the 
jet angle -- at X-ray energies (dominated by the receding beam).      

\subsection{Wolf-Rayet winds}

The winds from WRs are characterized by mass loss rates $\dot M\approx 
10^{-5}-10^{-4} \,\sfr$ and velocities $v_w\approx 1000-2500\,\kms$ (e.g. 
Willis 1991). In a steady, spherically symmetric wind, the electron density is
\begin{equation}
n_e(r)\approx 3\times 10^6\,{\rm 
cm}^{-3} \left({r\over 10^{15}\,{\rm cm}}\right)^{-2}\left({\dot M\over 
10^{-4}\,\sfr}\right) \left({v_w\over 1000\,\kms}\right)^{-1}
\mu_e^{-1}, \label{eq:ne}
\end{equation} 
where $\mu_e\sim 2$ in a helium gas. Figure 4 shows the fainter (compared 
to the red supergiant wind case) Compton echo of a GRB jet (with same 
parameters as above) propagating through a WR wind. 

In the case of a $r^{-2}$ medium with the fiducial scalings given
in equations (\ref{eq:rc}) and (\ref{eq:ne}), the electron scattering 
optical depth from $r_c$ to infinity is $\tau_c\approx 3\times 10^{-4}$. 
This is roughly the fraction of GRB energy which is reflected in the echo on 
all timescales $t\gta r_c/c\sim 6\,$ hours, with a 
scattered luminosity that drops as $t^{-2}$. Fast-moving winds will be less 
dense and therefore less efficient ($10^{-5}\lta \tau_c\lta 10^{-3}$) at 
reprocessing the prompt pulse; slow-moving winds will produce brighter echoes 
($10^{-4}\lta \tau_c\lta 0.05$). 
As the electron scattering optical depth is $\tau_c\propto r_c\propto 
(E/\Omega_b)^{-1/2}$, fainter and less collimated bursts will be characterized 
by a larger `albedo' relative to bright ones.
For an isotropic burst of energy (say) $E=10^{50}\,$ ergs, the characteristic 
distance in equation (\ref{eq:rc}) decreases to $r_c\approx 2\times 10^{13}\,$
cm. In a red supergiant wind with $n_e(r_c)\approx 4\times 10^{10}\,$
cm$^{-3}$, one has $\tau_c\approx 0.2$, 
and a fraction $(1-{\rm e}^{-\tau_c})\sim 20\%$ of the GRB energy will be 
reflected in a echo on timescales $\gta 10$ min.

\section{Discussion}
Bright scattering echoes are a natural consequence of a hard photon pulse 
propagating in a dense circumstellar environment such as a pre-burst stellar 
wind. In massive star progenitor models for GRBs there 
is likely to be an echo component in the observed X- and $\gamma$-ray light 
curves, the only question is how significant this component is. 
In the range 10 to 100 keV, where they mirror the spectral 
energy distribution of the prompt pulse, echoes will typically be harder than 
the afterglows observed by {\it BeppoSAX}. Above 200 keV, the reflected flash
will have a much steeper spectrum than the parent GRB as scattering 
occurs in the relativistic regime. While in the standard fireball/blastwave 
scenario both the prompt and delayed emission may be highly beamed, 
Compton echoes are -- modulo the scattering phase function -- quasi-isotropic.
Back-scattered radiation could then 
provide a means for detecting a population of nearby misaligned GRBs, since 
collimated outflows imply the existence of a large amount of undetected dim 
bursts and a much higher event rate than is often assumed. In a $\gamma$-ray
quiet burst, the observed luminosity of a Compton echo at fixed burst energy 
per steradian is proportional to the beaming solid angle, $L\propto\Omega_b$, 
as it scales with the intrinsic power of the parent GRB. In the Euclidean 
(bright) part of the number-flux relation the total number of echoes above a 
fixed limiting flux then scales as $L^{3/2}\Omega_b^{-1}
\propto \Omega_b^{1/2}$, i.e. it is smaller in the case of more collimated 
jets. In the flat (faint) part of the counts instead the number of echoes in 
the sky is approximately independent of the beaming solid angle.    

Compton echoes could be studied with the {\it Swift} Gamma Ray Burst Explorer
to be launched in $\sim 2003$. {\it Swift} will detect and follow GRBs with the 
Burst and Alert Telescope (BAT) at energies in the 10--100 keV range, together
with X-ray (XRT) and optical (UVOT) instrumentation (Gehrels 1999). Long 
duration $\gamma$-ray emission from the burst will be studied simultaneously 
with the X-ray and optical afterglow emission. With a sensitivity of 2 mCrab 
in a 16 h exposure, the BAT onboard of {\it Swift} will detect a $10^{45.7}\,
\lunits$ echo out to a distance of Gpc ($z\sim 0.2$).      

\subsection{Echoes versus X-ray afterglows}
It is interesting to compare the expected energetics of Compton echoes with the 
observed X-ray late afterglows. In a cosmology with $H_0=65\,\kmsmpc$, 
$\Omega_M=0.3$, and $\Omega_\Lambda=0.7$, the (isotropic equivalent) 
luminosity of GRB970228 three
days after the event was $\sim 4\times 10^{44}\,\lunits$ in the 2--10 keV band,
with a decay rate $\propto t^{-1.3}$ (Costa \etal 1997; Djorgovski \etal 
1999). 

Let us assume, for simplicity, that both the prompt and delayed GRB emission 
are {\it isotropic}. Had GRB970228 ($E\approx 10^{52}\,
$ergs) occurred in a SN 1993J-like environment (eq. \ref{eq:ne93J}), the 
expected light echo from an isotropic pulse, 
\begin{equation}
\epsilon L_\epsilon\approx 4\times 10^{44}\,\lunits\, \left({E\over 
10^{52}\,{\rm ergs}}\right)\,t_{\rm days}^{-1.5}
\left(\epsilon\over 10\,{\rm keV}\right) \label{eq:ec}            
\end{equation}
would be slightly fainter at 10 keV than the observed 
2--10 keV afterglow. At $t=10\,$h delay its 100 keV luminosity would exceed 
$10^{46}\,\lunits$, still consistent with the OSSE upper limit of Matz \etal  
(1997). A break in the echo light curve to a $t^{-2}$ decline
would be expected at $t\approx 2\times 10^{16}\, $cm$/c\sim 8\,$ day delay.   
Based on equation (\ref{eq:ec}), {\it we conclude that the scattered radiation 
from a pulse propagating in a red supergiant wind has a flux which is 
comparable to the  observed X-ray late afterglows.}
The echo would be about two orders of magnitude fainter in a WR-type wind on 
account of the greater speed.
Note that the light emitted during the early afterglow will also be scattered 
by circumstellar material and give rise to a light echo:
as the luminosity decays with time, however, `later' paraboloids will be 
sequentially dimmer, that is for any given radial distance from the 
burst the reflected power will actually decrease.

\subsection{X-ray delayed outbursts}
A delayed outburst was observed in the X-ray afterglow of GRB970508. 
The event had a luminosity of $\sim 2\times 10^{45}\,
\lunits$ (Piro \etal 1998; Metzger \etal 1997) at nine hour delay, with 
a decay $\propto t^{-1.1}$ up to $6\times 10^4\,$s. This was followed by a 
second flare of activity with a duration $\sim$ few $\times 10^5\,$s. The excess
energy was a significant fraction of the total, and the spectrum became harder 
during the flare (a possible detection of redshifted iron line emission has 
also been reported by Piro \etal 1999). The outburst could be explained by the 
Compton echo from a thin circumstellar shell of enhanced gas density in the 
neighborhood of the GRB. For an isotropic burst the reflected luminosity from 
the shell remains constant to within a factor of two [see Fig. 1, curve (b)].
While emission from the afterglow shock would then dominate over the 
scattered radiation at early times because of its steeper light curve,
this may not be necessarily true at later times.
For a shell radius of $R=1.5\times 10^{15}\,$cm 
(assumed to be larger than the distance reached by the shock producing the 
underlying afterglow emission), thickness $R/5$, and Thomson optical depth 
$\tau_T=0.2$, one derives a mean density $n_e\approx 5\times 10^8\,$cm$^{-3}$,
\footnote{This is a factor of ten higher than the characteristic density
of a red supergiant wind  at this distance (cf. eq. \ref{eq:ne93J}), an 
enhancement that could be caused by a variation of the mass-loss rate from 
the progenitor, by interacting slow and fast winds, by condensations formed 
via cooling instabilities, or by runaway $e^+e^-$ pair creation induced 
by collisions between soft side-scattered radiation and the main $\gamma$-ray
photon beam.}~ and a total mass $0.007\,M_\odot$ ($\mu_e=2$).
The echo reaches a maximum at $t_m=2R/c\approx 10^5\,$s, with luminosity  
\begin{equation}
\epsilon L_\epsilon={3 \tau_T \epsilon E_\epsilon \over 2t_m}\approx
7\times 10^{44}\,\lunits\, \left({E\over 10^{52}\,{\rm ergs}}\right)\,t_{m,\rm 
days}^{-1} \left(\epsilon\over 10\,{\rm keV}\right)            
\end{equation}
(cf. eq. \ref{eq:shell} and Fig. 1), enough to outshine the power-law decaying 
afterglow at late times. The spectrum in the flare would be harder as it 
mirrors the prompt emission. A temporary brightening should be observed 
again at later times as 
the afterglow shock reaches the shell and passes in front of it.         

\subsection{MeV delayed emission}
We have shown in \S\,3.2 (see also Figs. 2 and 3) that Compton downscattering
will produce a high-energy cutoff in the echo $\gamma$-ray emission at energies 
between $m_ec^2/2$ and $m_ec^2$.
There is a competing effect, however, which may generate MeV photons 
at late times. While in the standard fireball/blastwave scenario the source of 
X- and $\gamma$-ray radiation is itself expanding at relativistic speed and
photons are beamed into a narrow angle along the direction of motion, 
the light scattered off circumstellar material is quasi-isotropic and can 
interact with the relativistic ejecta via the bulk (inverse) Compton effect
(or Compton drag). In a $r^{-2}$ surrounding medium which is either at rest or
moving at sub-relativistic speed, the energy density $U$ of the reflected 
radiation drops as $r^{-4}$, and the effect will be dominated by the inner
regions close to the characteristic distance $r_c$. The scattered energy 
density at X-ray frequencies is
\begin{equation}
U_X={L_X\over 4\pi r_c^2c}={E_X n_e \sigma_T\over 4\pi r_c^2}
\end{equation}
(isotropic burst). 
An an illustrative possibility, consider a relativistic fireball made up by an 
individual shell of 
instantaneous  bulk Lorentz factor $\Gamma_F$, and let the scattering charges 
be cold in the fluid frame, $\langle \gamma_e\rangle \sim 1$. Seed photons of 
energy $\epsilon_X$ will then be upscattered to energies 
$\sim \epsilon_X\Gamma_F^2$ if $\epsilon_X\Gamma_F\ll m_ec^2$, or to 
$m_ec^2\Gamma_F$ otherwise. At $t=r_c/c\sim $ a few hours, the shock 
interaction of the relativistic ejecta with the circumstellar wind may have 
already slowed down the fireball to $\Gamma_F\sim$ a few. 
If $E$ is the total (initial) energy of the fireball, the instantaneous 
{\it emitted } power from bulk Compton scattering can be written as 
\begin{equation}
{\cal L}_e(\epsilon_X\Gamma_F^2)\approx \left({E\over m_p c^2\Gamma_F}\right) 
\left({\Gamma_F^2 E_X n_e \sigma_T^2c \over 4\pi r_c^2}\right), 
\end{equation}
where the first term in parenthesis is the number of particles being Compton
dragged at that instant (assuming the shock evolves adiabatically), and the 
second term is the rate of inverse Compton losses in the Thomson limit. 
The luminosity {\it received} at Earth is ${\cal L}_r\approx 2\Gamma_F^2{\cal 
L}_e$ due to the Doppler contraction of the observed time.
With $E=10^{53}\,$ergs, $\epsilon_X=40\,$keV, 
$E_X=4\times 10^{51}\,$ergs, $r_c=10^{15}\,$cm, $\Gamma_F=5$, and 
$n_e=10^{7}\,$ cm$^{-3}$, one derives ${\cal L}_r(1\,{\rm MeV}) \sim 7\times
10^{47}\,\lunits$ at an observed time $r_c/(2\Gamma_F^2c)\sim 10\,$ min. 
These numbers are only meant to be indicative, as they depend on the uncertain
evolution of the relativistic ejecta.
Note that, in the case of a burst propagating in a circumstellar wind, the 
pressure of the scattered photons will not be able to compete with the 
external material in braking the fireball, and therefore will not dictate 
its time evolution (see Lazzati \etal 2000 for a different scenario). For the 
opposite to be true the radiation energy density $U_X$ would have to exceed 
the rest-mass energy density of the scatterers. This would drive a 
relativistic outflow, suppressing the scattering rate and leading to an 
inconsistency.

\subsection{A lateral GRB jet from SN 1987A?}

It has been recently suggested by Cen (1999) that the bright, transient 
companion spot to SN 1987A observed about 1 month after the explosion 
(Nisenson \etal 1987; Meikle, Matcher, \& Morgan 1987)  may have been caused 
by a receding GRB jet traveling at $\theta=127^\circ$ with respect to the 
SN-to-observer direction, through a circumstellar medium with a stellar 
wind-like density $n_e\propto r^{-2}$. The scenario proposed by Cen has 
$n_e=1\,$cm$^{-3}$ at $r\approx 10^{19}\,$cm, an `isotropic equivalent' 
burst energy $4\pi E/\Omega_b=2\times 10^{54}\,$ ergs, and a beaming 
angle $\Omega_b=1.5\times 10^{-3}\,$rad. The late optical emission produced 
in an external shock model by synchrotron radiation appears then to provide 
an adequate explanation for the evolution of the observed companion spot. If 
the jet had approached us along the line of sight, a very bright GRB would 
have been observed instead.

With these parameters and the GRB spectrum given in (\ref{eq:spe}), one would
expect from equation (\ref{eq:Thom}) a hard X-ray echo of luminosity
\begin{equation}
\epsilon L_\epsilon\approx 10^{42}\,\lunits\, \left({E\over 10^{50}\,{\rm 
ergs}}\right) t_{\rm days}^{-2} \left({\epsilon\over 20\,{\rm keV}}\right).  
\end{equation}
From two to eight weeks after the explosion, however, no significant flux in 
the 10--30 keV band was observed by Ginga in the direction of 1987A to a crude 
upper limit of $10^{37}\,\lunits$ (Dotani \etal 1987; Makino 1987). The lack 
of a detectable Compton echo therefore places severe constraints on the 
brightness of a possible GRB jet associated with SN 1987A, as only an 
unusually weak burst of intrinsic energy $E\lta$ few$\times 10^{47}\,$ergs 
would be compatible with the assumed circumstellar density profile. 

\subsection{Optical echoes}
A strong prompt optical flash accompanied the brightest burst 
seen by {\it BeppoSAX}, GRB990123. The flash was observed by the Robotic
Optical Transient Search Experiment (ROTSE)
while the burst was still in progress, reached a peak of 9th magnitude, and 
then decayed with a power law slope of $-2$ (Akerlof \etal 1999). The redshift
of this burst ($z=1.6$, Kelson \etal 1999) implies a peak luminosity 
of $5\times 10^{49}\,\lunits$, and a total optical energy of $E_{\rm opt}
\approx 2\times 10^{51}\,$ ergs. An isotropic optical flash of this brightness,
occurring in a SN 1993J-like dense stellar wind, would give rise to an 
optical light echo of luminosity 
\begin{equation}
(\epsilon L_\epsilon)_{\rm opt}\approx 10^{45}\,\lunits\, \left({E_{\rm opt}
\over 2\times 10^{51}\, {\rm ergs}}\right) t_{\rm days}^{-1.5}.  
\end{equation}
Here we have assumed that the opacity is dominated by electron scattering, as 
the prompt flash will photoionize the ambient medium and destroy any dust by 
thermal sublimation out to a radius $\sim 1\,$pc (Waxman \& Draine 1999).
However, beyond this radius, the refractory cores of dust grains 
can survive until they are passed by the expanding blast wave. These
grains have high albedo, selective extinction and forward
scattering and may scatter the GRB light from the first few hours to form a 
supernova-like optical echo after a few months (Esin \& Blandford 2000).

Two days after the event the transient afterglow was observed in the $r$-band
at a level of $7\,\mu$Jy (Kulkarni \etal 1999), or $(\epsilon L_\epsilon)_r
\approx 8\times 10^{43}\,\lunits$, quite a bit fainter than the expected 
optical echo. The data would then appear to rule out a dense red supergiant 
environment for GRB990123 unless the prompt optical flash is actually beamed,
which it may be.  

\acknowledgments
We have benefited from many useful discussions with G. Ghisellini, D. Helfand, 
E. Ramirez-Ruiz, and C. Thompson. Support for this work was provided by NSF 
through grant PHY94-07194 (P. M. and R. D. B.), by NASA through grant 5-2837 
and the Beverly and Raymond Sackler Foundation (R. D. B.), and by the Royal 
Society (M. J. R.).

\references

Akerlof, C., \etal 1999, Nature, 398, 400

Esin, A. A., \& Blandford, R. D. 2000, ApJ, in press

Blandford, R. D., \& Rees, M. J. 1972, Astrophys. Lett., 10, 77

Bloom, J. S., \etal 1999, Nature, 401, 453 

Cen, R. 1999, ApJ, 524, L51

Chevalier, R. A., \& Li, Z.-Y. 1999, ApJ, 520, L29

Costa, E., \etal 1997, Nature, 387, 783

Djorgovski, S. G., \etal 1999, GCN Report 289 

Dotani, T., \etal 1987, Nature, 330, 230

Fransson, C., Lundquist, P., \& Chevalier, R. A. 1996, ApJ, 461, 993

Fruchter, A. S., \etal 1999, ApJ, 520, 54 

Galama, T. J., \etal 1998, Nature, 395, 670

Gehrels, N. 1999, BAAS, 31, 993

Harrison, F. A., \etal 1999, ApJ, 523, L121

Kelson, D. D., Illingworth, G. D., Franx, M., Magee, D., van Dokkum, 
P. G. 1999, IAUC 7096

Kulkarni, S. R., \etal 1999, Nature, 398, 389

Lazzati, D., Ghisellini, G., Celotti, A., \& Rees, M. J. 2000, ApJ, 529, L17

MacFadyen, A., \& Woosley, S. E. 1999, ApJ, 524, 262

Madau, P., \& Thompson, C. 2000, ApJ, in press 

Makino, F. 1987, IAUC 4336

Matz, S. M., McNaron-Brown, K., Grove, J. E., Share, G. H. 1997, IAUC 6678

Meikle, W. P. S., Matcher, S. J., \& Morgan, B. L. 1987, Nature, 329, 608

M\'{e}sz\'{a}ros, P., Rees, M. J. 1992, ApJ, 397, 570

M\'{e}sz\'{a}ros, P., Rees, M. J., \& Wijers, R. A. M. J. 1999, NewA, 4, 303 

Metzger, M. R., \etal 1997, IAUC 6655

Moderski, R., Sikora, M., \& Bulik, T. 2000, ApJ, 529, 151

Nisenson, P., Papaliolios, C., Karovska, M., \& Noyes, R. 1987, ApJ, 320, L15

Paczy\'{n}ski, B. 1998, ApJ, 494, L45

Paczy\'{n}ski, B. 1999, in The Largest Explosions Since the Big Bang:
Supernovae and Gamma-Ray Bursts, eds. M. Livio, K. Sahu, and N. Panagia
(Cambridge: Cambridge University Press), in press 

Piran, T. 1999, Phys. Rep., 314, 575

Piro, L., \etal 1998, A\&A, 331, L41

Piro, L., \etal 1999, ApJ, 514, L73
 
Preece, R. D., \etal 2000, ApJS, 126, 19 

Reichart, D. E. 1999, ApJ, 521, L111 

Rybicki, G.~B., \& Lightman, A.~P. 1979, Radiative Processes in Astrophysics
(New York: Wiley)

Thompson, C., \& Madau, P. 2000, ApJ, in press

Vietri, M., \& Stella, L. 1998, ApJ, 507, L45


Waxman, E., \& Draine, B. T. 1999, ApJ, submitted (astro-ph/9909020)

Willis, A. J. 1991, in Wolf-Rayet Stars and Interrelations with Other
Massive Stars in Galaxies, ed. K. A. van der Hucht \& B. Hidayat
(Dordrecht: Kluwer), 256

van Paradijs, J., \etal 1997, Nature, 386, 686
\page

\begin{figure}
\plotone{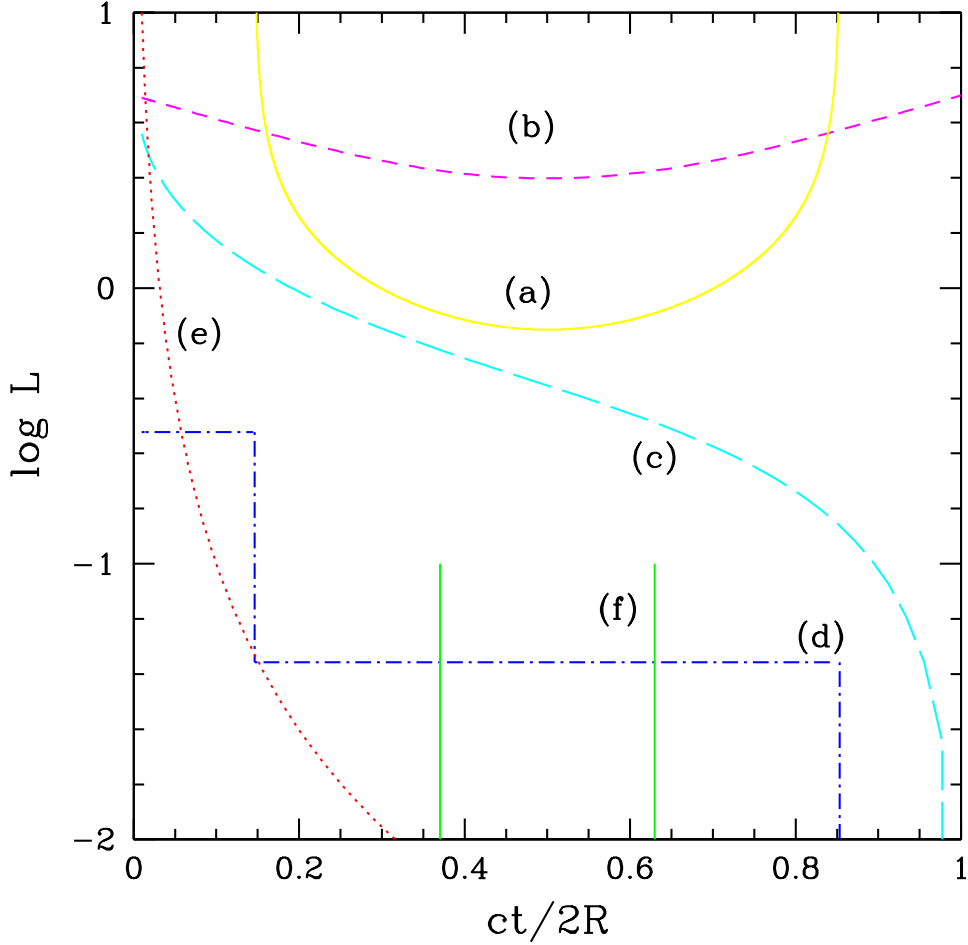}
\caption{\small Possible X-ray flux variations observed following a GRB, and due to 
Thomson scattering in the circumstellar medium. A variety of ambient gas 
distributions has been assumed (see text for details): the scattered fluxes, 
espressed as equivalent isotropic luminosities, have arbitrary normalizations. 
{\it Solid curve (a)}: slender annular ring of radius $R$, inclined at 
$45^\circ$ to the line of sight, isotropic burst. {\it Short-dashed line (b)}: 
thin spherical shell of radius $R$, isotropic burst. {\it Long-dashed line 
(c)}: uniform density sphere of radius $R$, isotropic burst. 
{\it Dash-dotted line (d)}: uniform density sphere of radius $R$, collimated
burst inclined at $45^\circ$ to the line of sight. {\it Dotted line (e)}: 
constant velocity wind ($n_e\propto r^{-2}$). 
{\it Double spike (f)}: thin spherical shell of radius $R$, burst is collimated
into two anti-parallel beams inclined at $75^\circ$ to the line of sight. 
\label{fig1}}
\end{figure}
\vfill\eject

\begin{figure}
\plotone{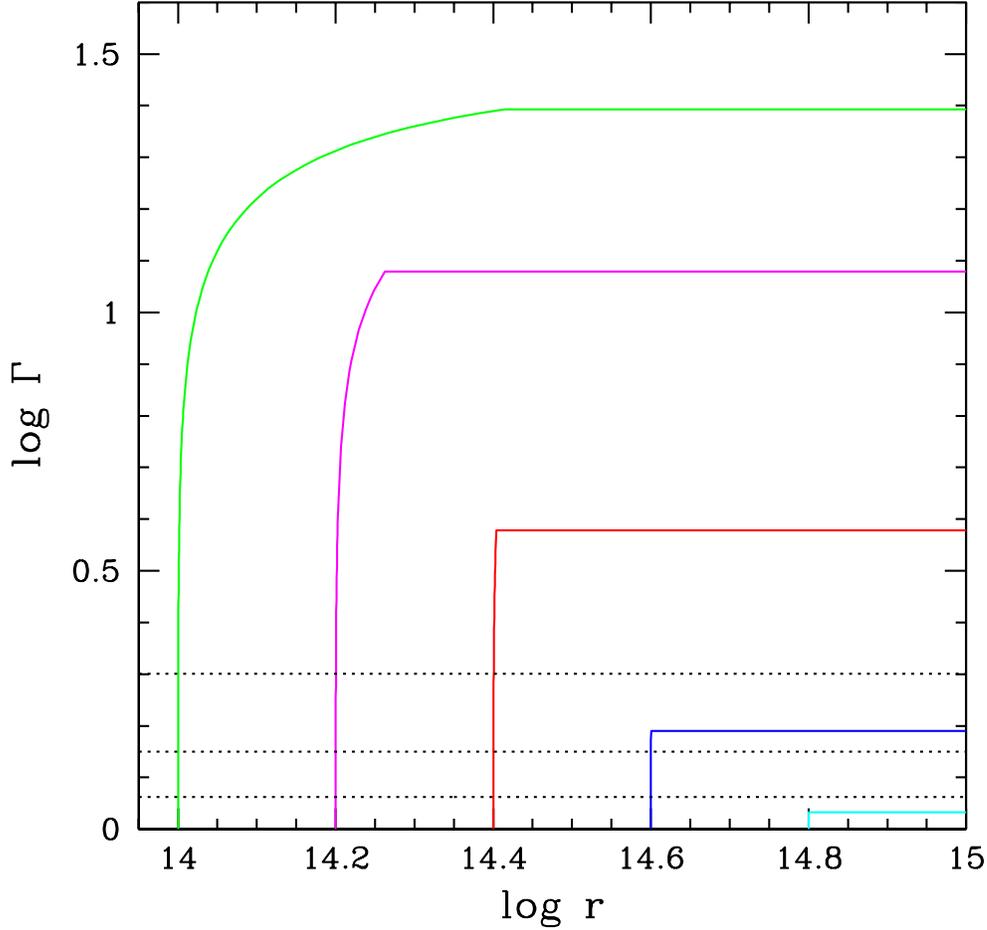}
\caption{\small Bulk Lorentz factor of optically thin, circumstellar material 
with molecular weight per electron $\mu_e=2$, as a function of Lagrangian 
distance $r$ (in cm) from a collimated source of impulsive radiation. 
The photon pulse (assumed to be plane-parallel) has an `isotropic-equivalent' 
energy of $10^{53}\,$ergs, duration $10\,$s, and spectrum as in (\ref{eq:spe}). 
The radiative force vanishes when the photon shell moves past the particle.
The equation of motion has been integrated in the Klein-Nishina regime assuming 
the material to be initially at rest at $\log r \,({\rm cm})=14, 14.2, 14.4, 
14.6,$ and 14.8 ({\it solid curves}). The accelerated medium will be compressed
into a shell of thickness $\sim r/\Gamma^2$. Shocks may form when 
inner shells (which move faster and are more compressed) run into outer shells,
and material will accumulate at a radius $r_c\sim 6\times 10^{14}\,$cm where 
the outflow becomes sub-relativistic. The three dotted lines show the value of 
$\Gamma$ for which $\beta=\cos\theta$, where $\theta=(30^\circ, 45^\circ, 
60^\circ)$ is the angle between the line of sight and the approaching jet of
a collimated burst. Above these lines the scattered radiation will be `beamed
away' from the line of sight.    
\label{fig2}}
\end{figure}
\page

\begin{figure}
\plottwo{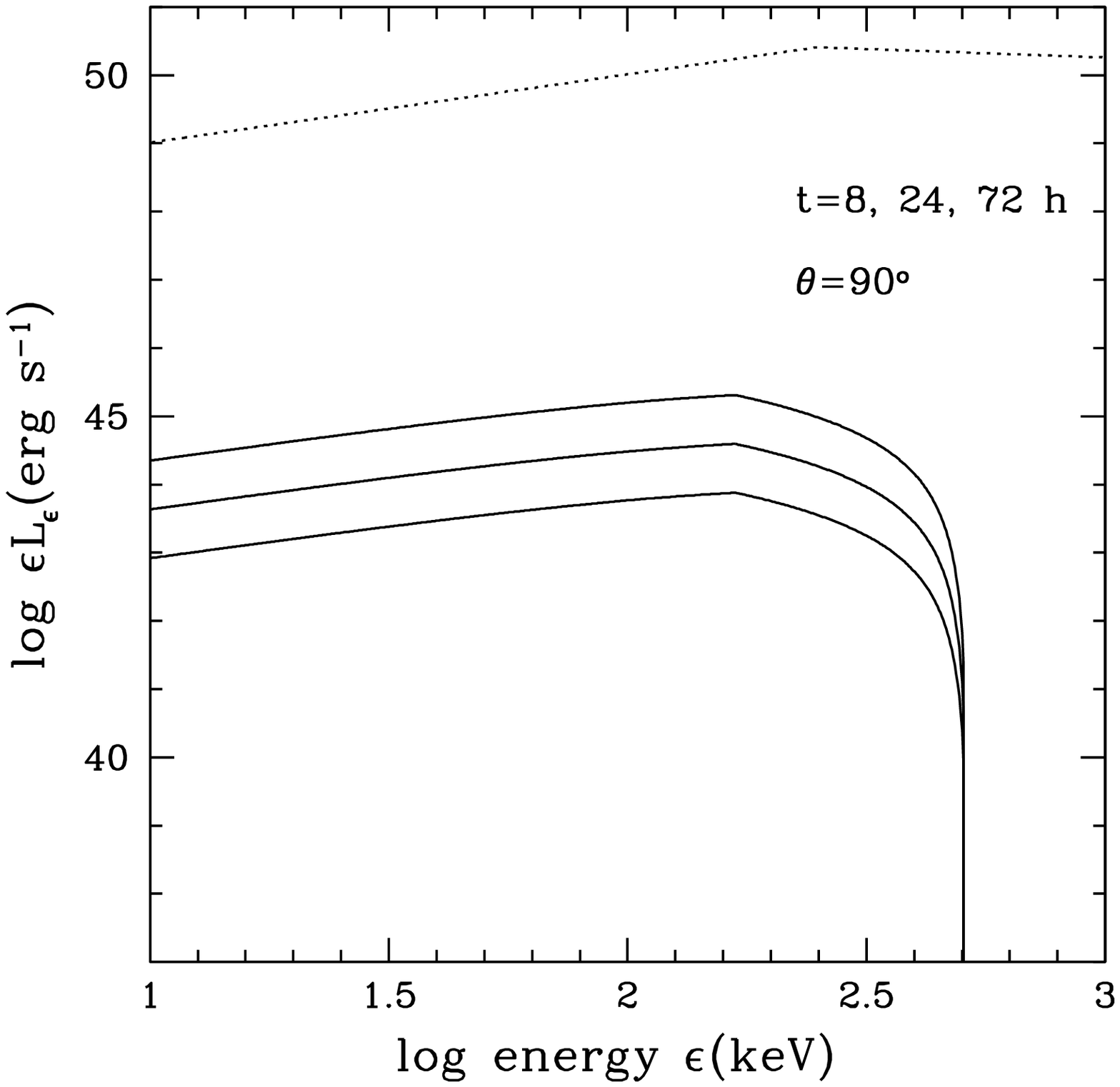}{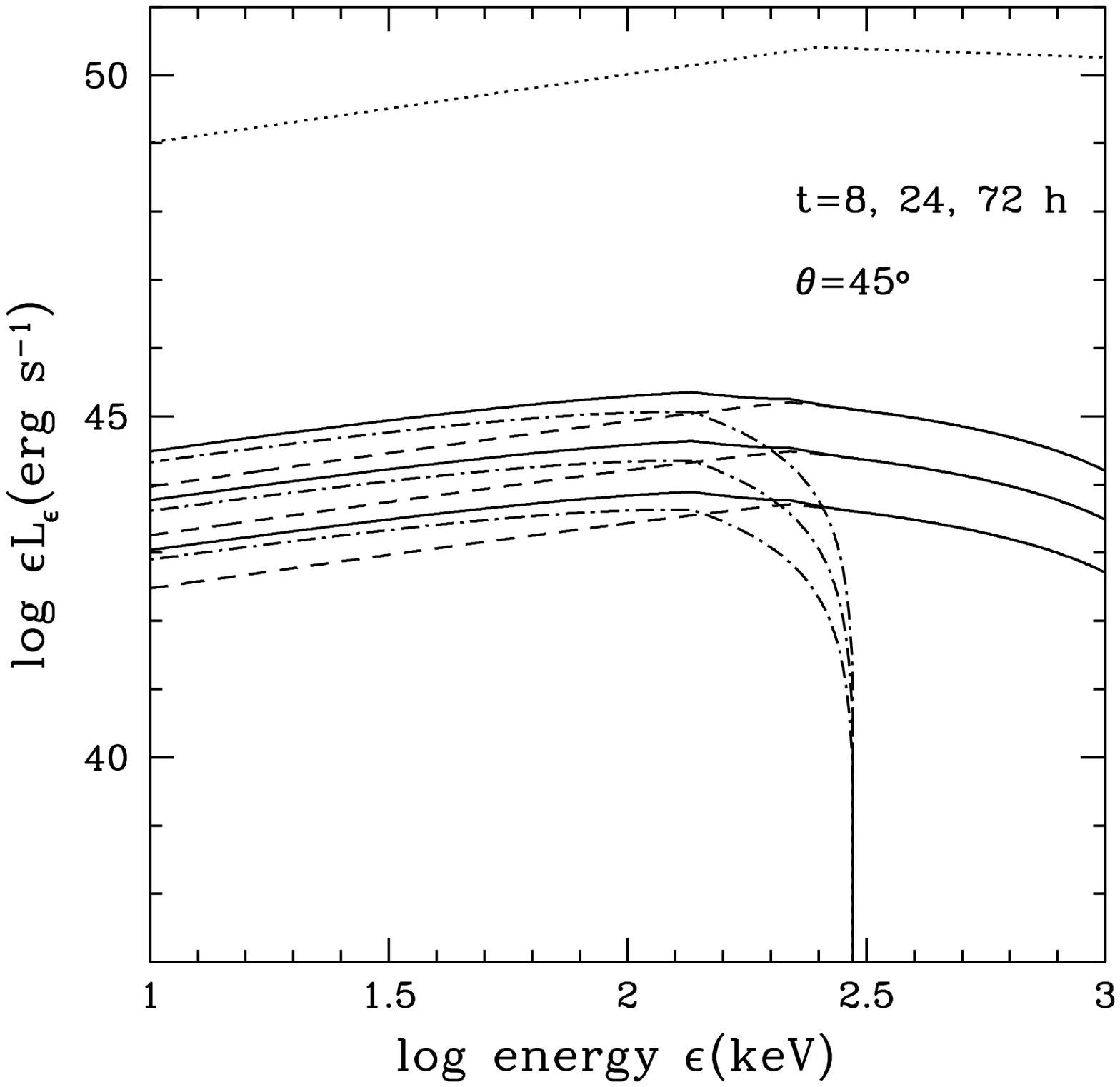}
\caption{\small The Compton echo of a GRB. The primary burst is assumed to be 
a two-sided collimated pulse of total energy $10^{52}\,$ ergs and duration 10 s
({\it dotted} spectrum at the top of each panel), propagating at an angle 
$\theta$ with the line of sight, and invisible to the observer. The 
circumstellar red supergiant
wind has a $r^{-1.5}$ density profile up to $2\times 10^{16}\,$cm, 
steepening to a $-2$ slope at larger radii. The electron density is normalized
to $n_e=10^8\,$ cm$^{-3}$ at $r=10^{15}\,$cm. 
{\it Left:} $\theta=90^\circ$. {\it Solid lines:} 
observed echo at $t=8, 24$ and 72 hours after the burst (from top to bottom). 
{\it Right:} $\theta=45^\circ$. {\it Solid lines:} Same as before.
{\it Dashed lines:} reflected light from approaching beam.
{\it Dash-dotted lines:} reflected light from receding beam.
\label{fig3}}
\vspace{+0.5cm}
\end{figure}
\page

\begin{figure}
\plottwo{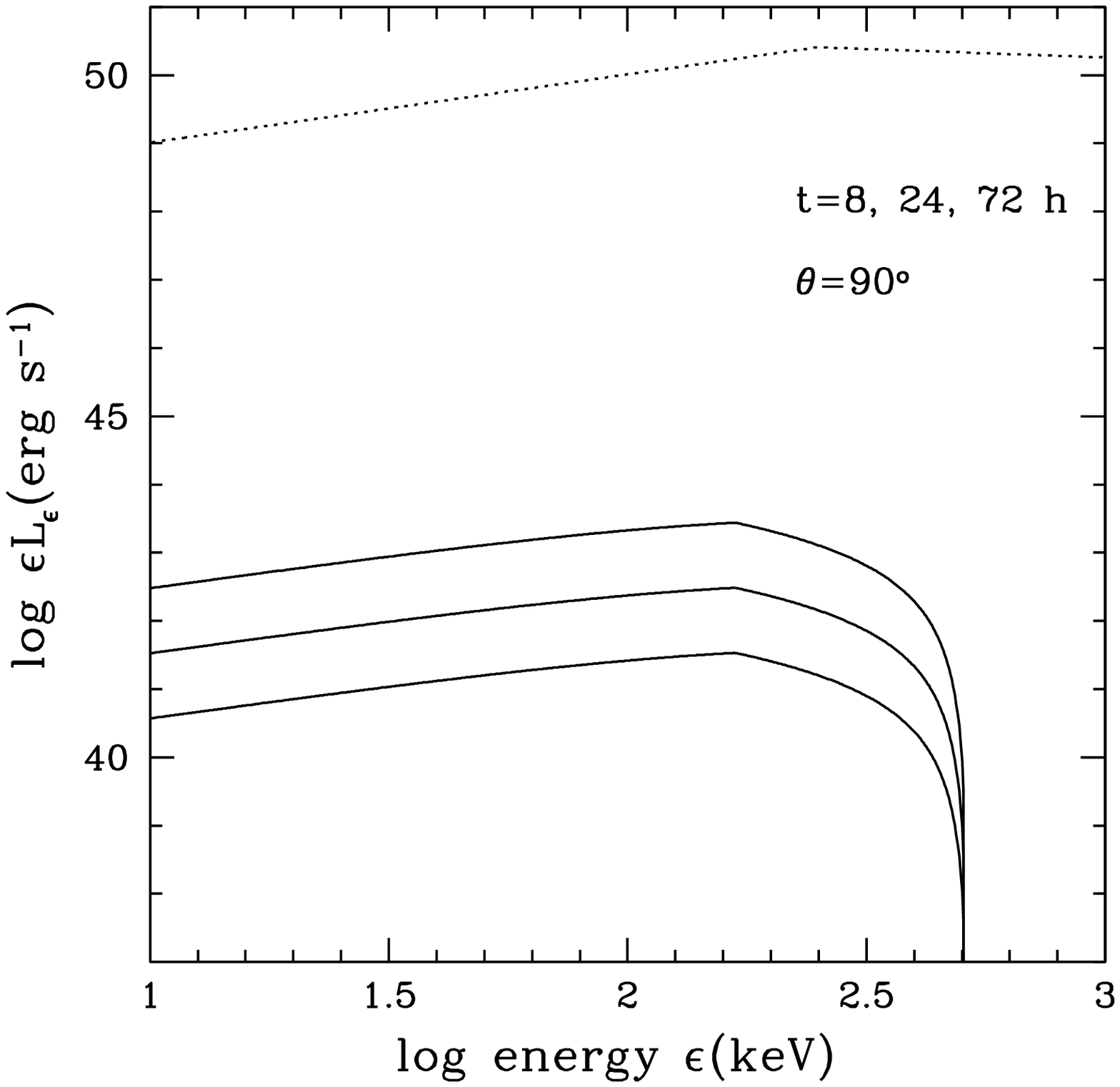}{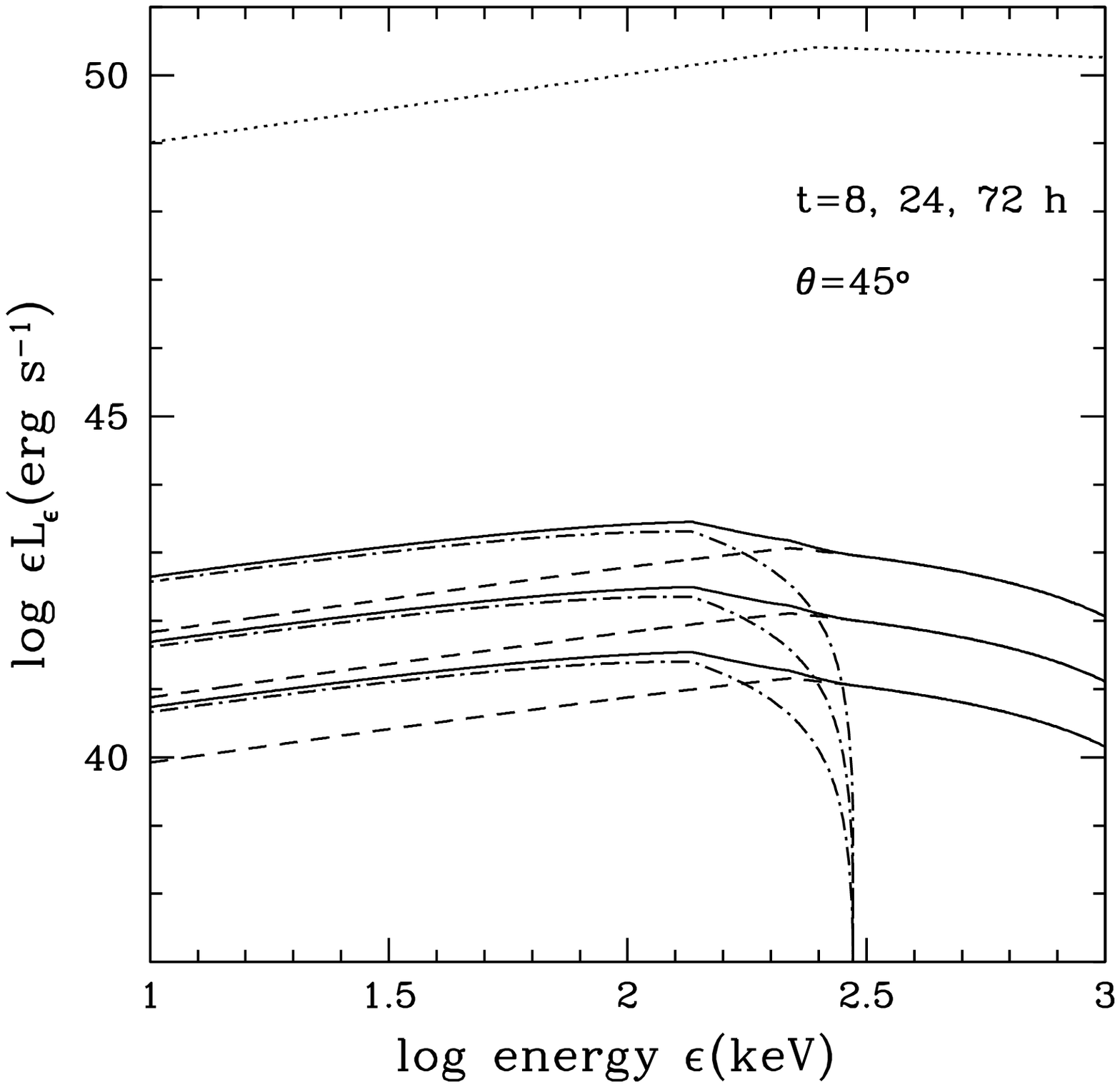}
\caption{\small Same as Figure 3, except that the circumstellar WR wind has a $r^{-2}$
density profile with electron density $n_e=1.5\times 10^6\,$cm$^{-3}$ at 
$r=10^{15}\,$ cm. The scattered luminosity drops as $t^{-2}$ in all cases.
\label{fig4}}
\end{figure}

\end{document}